\documentclass[11pt,aps,nofootinbib,showpacs,longbibliography,preprintnumbers,
superscriptaddress,floatfix,notitlepage]{revtex4-1}
\usepackage{amsmath,amsfonts}
\usepackage{graphicx}

\def\be{\begin{equation}}
\def\ee{\end{equation}}
\def\bea{\begin{eqnarray}}
\def\eea{\end{eqnarray}}

\begin{document}

\preprint{TIFR/TH/11-02}

\title{Top quark forward-backward asymmetry at Tevatron and its implications at the LHC }

\author{Biplob Bhattacherjee}
\email{biplob@theory.tifr.res.in}
\affiliation{Tata Institute of Fundamental Research, 
Homi Bhabha Road, Colaba, Mumbai 400005, India}
\author{Sudhansu S. Biswal}
\email{sudhansu@theory.tifr.res.in}
\affiliation{Tata Institute of Fundamental Research, 
Homi Bhabha Road, Colaba, Mumbai 400005, India}
\author{Diptimoy Ghosh}
\email{diptimoyghosh@theory.tifr.res.in}
\affiliation{Tata Institute of Fundamental Research, 
Homi Bhabha Road, Colaba, Mumbai 400005, India}


\begin{abstract}
The measurement of forward-backward asymmetry in the top and anti-top quark ($t \bar{t}$)
production has been recently reconfirmed by the CDF Collaboration and shows a more 
than $3\sigma$ deviation from the Standard Model(SM) prediction in the large $t \bar{t}$ 
invariant mass region.
Models with new W$^{\prime}$ or 
Z$^{\prime}$ bosons have been invoked to explain this deviation.
In the context of these models we perform a $\chi^2$ analysis with all the available 
experimental numbers in different $\Delta Y$ and $M_{t \bar{t}}$ bins. 
We show that for the $Z^{\prime}$ model the region of parameter space which explain 
the Tevatron asymmetry can be probed in the same sign top production channel by Tevatron 
itself. Moreover, we consider a recently proposed observable, the one sided forward-backward 
asymmetry ($A_{OFB}$) at the LHC and conclude that both the W$^{\prime}$ and Z$^{\prime}$ models 
can lead to sizable $A_{OFB}$ even at the LHC running at a center of mass energy of 7 TeV 
for the model parameters consistent with the Tevatron measurements.  
\end{abstract}

\maketitle

\section{Introduction}
The top quark with its mass close to the electroweak symmetry breaking scale and 
being about 40 times heavier than the next heavy quark is expected to be crucially 
sensitive to the physics which underlie the mechanism of electroweak symmetry 
breaking. 
Many properties of top quark have been undergoing serious examination at the Fermilab 
Tevatron and LHC, being a top factory, will study the properties of top quark with 
unprecedented precision.
The forward-backward(FB) asymmetry of top quark pairs $A_{FB}^{t \bar{t}}$ in $p \,\bar{p}$ 
collisions was measured by Tevatron with $\sqrt{s}$=1.96 TeV in 2008, which is defined as:
\be
A_{FB}^{t \bar{t}} \equiv 
\frac{\sigma(\Delta Y > 0)-\sigma(\Delta Y < 0)}{\sigma(\Delta Y > 0)+\sigma(\Delta Y < 0)} \; ,
\label{eq.1}
\ee
where $\Delta Y \equiv Y_t -Y_{\bar{t}}$ , the difference of rapidities of the top and 
anti-top quarks respectively in an event.   
The D$\O{}$ collaboration\cite{D0:2007qb} measured $[12\pm 8(\text{stat}) \pm 1(\text{sys})]\%$ 
asymmetry with 0.9 fb$^{-1}$ data for $t \bar{t} + X$ events with four or more jets while the CDF 
collaboration\cite{Aaltonen:2008hc} reported $[24 \pm 13(\text{stat}) \pm 4 (\text{sys})]\%$ 
parton level asymmetry with 1.9 fb$^{-1}$ data.

In the SM the FB asymmetry arises at the order $\alpha_s^3$ \cite{Bowen:2005ap,
Antunano:2007da,Almeida:2008ug} in QCD from i) interference between the tree level amplitude 
and the box diagram, ii) initial and final state gluon bremsstrahlung, iii) 
gluon-quark annihilation and scattering into $t \bar{t}$ final state. 
The size of this asymmetry is predicted to be $[6 \pm 1]\%$\cite{Aaltonen:2011kc} in the SM.
Though the SM prediction is consistent with the experimental numbers within 2$\sigma$, the 
large central value of the asymmetry has provoked theorists to propose possible new 
physics scenarios\cite{Jung:2009jz,Cheung:2009ch,Shu:2009xf,Cao:2009uz,Ferrario:2009bz,
Arhrib:2009hu,Dorsner:2009mq,Jung:2009pi,Cao:2010zb,Xiao:2010hm,Barger:2010mw,
Bauer:2010iq,Chen:2010hm,Choudhury:2010cd,Chivukula:2010fk,Degrande:2010kt,Cao:2010nw,Jung:2010yn,
Alvarez:2010js,Delaunay:2011vv,Cao:2011ew,Bai:2011ed} which can give rise to large forward-backward 
asymmetry.
Recently the CDF collaboration has updated their result with much more data of 5.3 fb$^{-1}$ 
to get the parton level total asymmetry $ A_{FB}^{t \bar{t}} = 0.158 \pm 0.075(\text{stat+syst})$ 
\cite{Aaltonen:2011kc} which reconfirmed their earlier measurement.
More interestingly, the forward-backward asymmetry is observed to be more pronounced in the large 
$t\,\bar{t}$ invariant mass region and in the region where the rapidity difference $\Delta Y$ is 
large. We quote their results in Table-\ref{Table1} for better readability. 
\begin{table}[h]
$$
\begin{array}{|c|c|c|}
\hline
\text{Observable} &A_{FB}^{t \bar{t}}(|\Delta Y|<1.0) & A_{FB}^{t \bar{t}}(|\Delta Y|>1.0) 
\\ \hline 
\text{CDF result}        & 0.026\pm 0.118 & 0.611\pm 0.256    \\ \hline
\text{SM Prediction}& 0.039\pm 0.006 & 0.123\pm 0.008  \\ \hline
\text{Observable}& A_{FB}^{t \bar{t}}(M_{t \bar{t}} < 450 ~GeV) & A_{FB}^{t \bar{t}}(M_{t \bar{t}} > 450 ~GeV)
\\ \hline
\text{CDF result}        & -0.116\pm 0.153 & 0.475\pm 0.114  \\ \hline
\text{SM Prediction}& 0.040\pm 0.006 & 0.088\pm 0.013  \\ \hline
\end{array}
$$
\caption{CDF measurements and SM predictions of the Forward-Backward Asymmetry in different 
$\Delta Y$ and $M_{t \bar{t}}$ bins.}
\label{Table1}
\end{table}
From Table-\ref{Table1} one should notice that the asymmetry at high invariant  
mass region is more than 3 standard deviations above the NLO SM prediction.
It is intriguing that though the forward backward asymmetry 
shows a clear deviation from SM QCD prediction at least in the large 
$t \bar{t}$ invariant mass region, the measured parton level $t \bar{t}$ 
cross section $\sigma^{\text{Measured}}_{t \bar{t}} =7.70 \pm 0.52$ pb 
\cite{Aaltonen:2010ic} and invariant mass distribution\cite{Aaltonen:2009iz} 
are still consistent with SM 
prediction $\sigma^{\text{SM}}_{t \bar{t}}(\text{MCFM}) =7.45^{+0.72}_{-0.63}$ pb \cite{Moch:2008ai}.
Hence, any model which will explain the invariant mass dependent asymmetry should 
also accommodate the observed consistency of the invariant mass distribution with SM.
To this end, we consider the new physics scenarios with a t-channel vector boson exchange such
as a new flavor changing $Z^{\prime}$\cite{Jung:2009jz} or a new $W^{\prime}$\cite{Cheung:2009ch}. 
A s-channel vector boson mediated $q \bar{q} \to t \bar{t}$ process can also produce the required 
asymmetry, but also increases the $t \bar{t}$ production cross section which is measured to be 
consistent with SM \cite{Sehgal:1987wi,Bagger:1987fz,Choudhury:2007ux,Frampton:2009rk,Djouadi:2009nb,
Bauer:2010iq}. Unlike s-channel 
exchange, a t-channel diagram generally has the advantage of not changing the cross section appreciably. 
Apart from generating the forward-backward asymmetry the t-channel $Z^{\prime}$ scenario also 
contributes to same sign top production, single top production and FCNC top quark decays which make 
this model very interesting. On the other hand, the $W^{\prime}$ model has no same sign top signal 
and it is challenging to see how this model can be probed at Tevatron or LHC.

This paper is organized as follows. In the next section we briefly describe the Z$^{\prime}$ and 
W$^{\prime}$ models and perform a $\chi^2$ analysis of their parameter spaces and study some 
of their collider signatures. In section ~\ref{section3} we consider a recently proposed observable 
called the one sided forward-backward asymmetry ($A_{OFB}$) \cite{Wang:2010du,Wang:2010tg} and calculate 
it at 7 TeV LHC for both these models. We discuss our results and summarize in section ~\ref{section4}. 
\section{Scenario with a new Z$^{\prime}$/W$^{\prime}$ boson}
\label{section2}
We parametrize the Lagrangian for the Z$^{\prime}$ model as  
\be
\mathcal{L} \ni g_{_{Z^{\prime}}} \bar{u} \gamma^{\mu} P_R t \, Z^{\prime}_{\mu} + 
\epsilon_{_U} g_{_{Z^{\prime}}} \bar{u_i} \gamma^{\mu} P_R u_i \, Z^{\prime}_{\mu} + h.c.\;,
\ee
where $g_{_{Z^{\prime}}}$, $\epsilon_{_U}$ are the new coupling constants and $i$ is the generation 
index. 
In this analysis we do not consider new (V-A) couplings as they are highly restricted from the 
$B_d-\bar{B_d}$ mixing measurements\cite{Berger:2011ua}. 

Note that the new  $Z^{\prime}$ contributes to both the single top production via 
$u g \to t Z^{\prime}(\to u_i \bar{u_i})$ as well as the same sign top production via t-channel 
$u(\bar{u})u(\bar{u}) \to t(\bar{t})t(\bar{t})$, $u(\bar{u}) g \to t(\bar{t}) 
Z^{\prime}(\to t(\bar{t}) \bar{u}(u))$ 
and  $u\bar{u} \to Z^{\prime}(\to \bar{u} t) Z^{\prime}(\to \bar{u} t)$ processes.
The term proportional to  $\epsilon_{_U}$ give rise to the decay modes 
$Z^{\prime} \to \bar{u_i} u_i$. If mass of $Z^{\prime}$ is greater than the top quark mass 
then this helps reducing the same sign top quark production via $u\bar{u} \to 
Z^{\prime}(\to \bar{u} t) Z^{\prime}(\to \bar{u} t)$ and $u(\bar{u}) g \to t(\bar{t}) 
Z^{\prime}(\to t(\bar{t}) \bar{u}(u))$ . 

We now consider the six measured observables 
$\sigma^{t \bar{t}}(\text{total}), A_{FB}^{t \bar{t}}(\text{total}), 
A_{FB}^{t \bar{t}}(M_{t \bar{t}}< 450 \rm{GeV})$, $ A_{FB}^{t \bar{t}}(M_{t \bar{t}} > 450\rm{GeV}), 
A_{FB}^{t \bar{t}}(|\Delta Y|<1.0), A_{FB}^{t \bar{t}}(|\Delta Y|>1.0)$ and try to find out 
the favoured parameter space of the $Z^{'}$ model. To do this we define the $\chi^2$ function
as: 
\be
\chi^2= \sum_i\frac{(O_i^{\text{Theory}} - O_i^{\text{Measured}})^2}{\sigma_i^2} ,
\ee
where $O_i$ are the six observables. We add the 
experimental and Standard Model errors in quadrature to calculate $\sigma_i$. 

For numerical studies we use $m_t=172.5$ GeV. 
To get the correct SM  $t \bar{t}$ production cross section at Tevatron we use the QCD K 
factor=1.3. We set both the renormalization and factorization scales to be $m_t$ and 
convolute the parton level cross section with CTEQ6L parton distribution functions. 
We use CalcHEP \cite{Pukhov:2004ca} for parton level analysis. 
Note that our $O_i^{\text{Theory}}$ includes both 
the new physics and the Standard Model contributions.
 
\begin{figure}[h]
\centering
\hspace*{0mm}\includegraphics[width=0.75\textwidth]{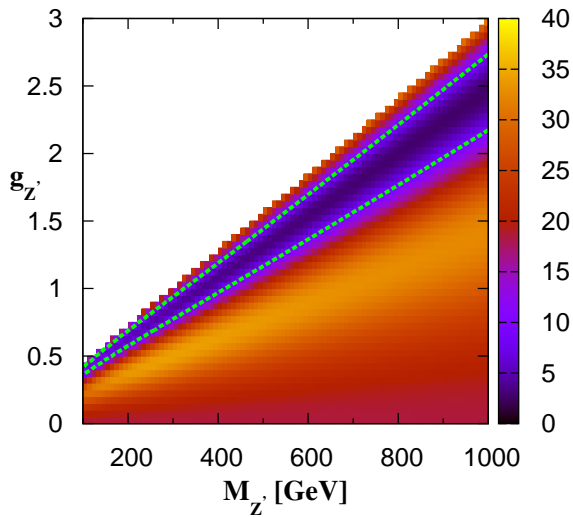}
\caption{
$\chi^2$ distribution in the $M_{_{Z^{\prime}}}- g_{_{Z^{\prime}}}$ plane for the 
$Z^{\prime}$ model. The area between the two dashed lines corresponds to the 99\% CL region.}
\label{Fig.1}
\end{figure}

\begin{figure}[h]
\centering
\hspace*{0mm}\includegraphics[width=0.6\textwidth]{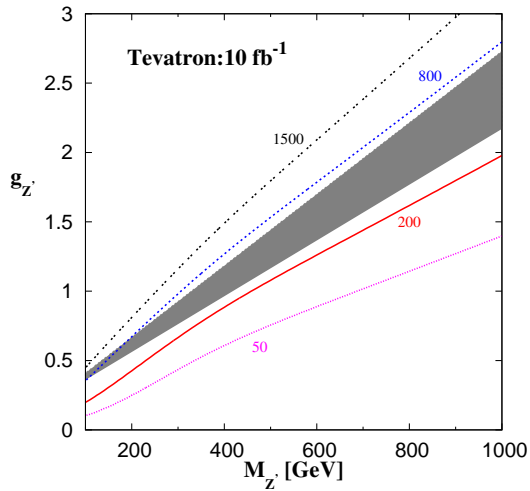}
\caption{
Number of same sign dilepton events from the same sign top pair decays at Tevatron. 
The shaded area is the 99\% CL allowed region of Fig.\ref{Fig.1}.}
\label{Fig.2a}
\end{figure}

\begin{figure}[h]
\centering
\hspace*{0mm}\includegraphics[width=0.6\textwidth]{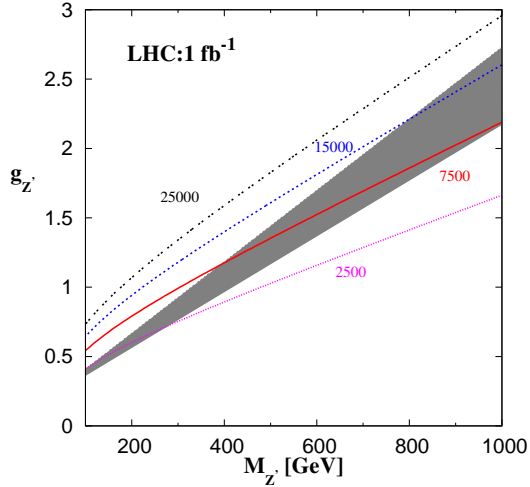}
\caption{
Number of same sign dilepton events from the same sign top pair decays at 7 TeV LHC. 
The shaded area is the 99\% CL allowed region of Fig.\ref{Fig.1}.}
\label{Fig.2b}
\end{figure}

Fig.\ref{Fig.1} shows the $\chi^2$ distribution in the $M_{_{Z^{\prime}}}- g_{_{Z^{\prime}}}$ 
plane for the $Z^{\prime}$ model. The region between the two dashed lines corresponds to 
99\% confidence level which is obtained by the frequentist approach assuming all errors to
be Gaussian. Notice that large values of the $Z^{\prime}$ 
mass, consistent with the Tevatron data, are also possible if one allows for large coupling. 
Note that the uncoloured 
region in Fig.~\ref{Fig.1} has $\chi^2$ more than 40, primarily because of very large 
$t \bar{t}$ production cross section.
As mentioned before, the existence of the $Z^{\prime}$ boson contributes to 
the same sign top pair production via $u(\bar{u})u(\bar{u}) \to t(\bar{t})t(\bar{t})$, 
$u(\bar{u}) g \to t(\bar{t}) Z^{\prime}(\to t(\bar{t}) \bar{u}(u))$ 
and  $u\bar{u} \to Z^{\prime}(\to \bar{u} t) Z^{\prime}(\to \bar{u} t)$ channels. 
The second and the third channels contribute only if the mass of $Z^{\prime}$ is greater than the top 
quark mass. These two contributions can be decreased by increasing the coupling  $\epsilon_U$, though 
very large value of $\epsilon_U$ may contradict with the di-jet resonance search at both 
the Tevatron and LHC \cite{Aaltonen:2008dn,Khachatryan:2010jd}. 
On the other hand the first channel is independent of $\epsilon_U$. 
We consider $\epsilon_U$ in the range $\sim$ $10^{-2} - 10^{-1}$ for the numerical analysis. 
All our results are practically independent of the value of $\epsilon_U$ for the above range. 
The leptonic branching ratio of top quark is 
about 0.22 (considering electron and muon only). Thus, about 5\% of the same sign top quark pair 
decays through the same sign dilepton channel. In Fig.\ref{Fig.2a} we show the number of 
same sign dilepton events from the same sign top pair decays  
expected at Tevatron with 10fb$^{-1}$ data. 
As an example, for 
$M_{_{Z^{\prime}}} = 200$ GeV and $g_{_{Z^{\prime}}}=0.6$ the total same sign top production 
cross section in the three channels mentioned at Tevatron is about 1.1 pb giving rise to 
about 550 same sign dilepton 
events at 10 fb$^{-1}$. 
CDF has searched for like sign dilepton events plus $b$ jet and missing 
transverse energy and found only 3 such events with 2fb$^{-1}$ of data \cite{Aaltonen:2008hx} 
which is consistent with the SM expectation. We see that the parameter values which 
explain the Tevatron asymmetry quite well also predict quite a few same sign top pairs at the 
Tevatron with 10$fb^{-1}$ of data. Note that, in real 
experiment, the number will be much smaller than our numbers because of detector effects. 
Following the cuts mentioed in \cite{Aaltonen:2008hx} we get roughly about 20\% efficiency 
(including b tagging) in a PYTHIA \cite{Sjostrand:2006za} level simulation. 
Hence, out of the 550 same sign dilepton events mentioed above we expect about 100 events to survive the 
standard experimental cuts. A detailed study of all the backgrounds and detector effects is beyond the scope 
of this work. Nevertheless, we expect a 
large part of the parameter space in the $M_{_{Z^{\prime}}}- g_{_{Z^{\prime}}}$ plane can 
already be probed with the collected data at Tevatron. 
At the LHC the situation is much better than 
Tevatron \cite{Cao:2011ew,Berger:2011ua} as can be seen in the Fig.\ref{Fig.2b}.
Fig.\ref{Fig.2b} is similar to Fig.\ref{Fig.2a}, but with a smaller 
integrated luminosity of 1$fb^{-1}$ at 7 TeV LHC and the numbers are only for the 
$u(\bar{u})u(\bar{u}) \to t(\bar{t})t(\bar{t})$ channel. To compare with Tevatron, for the same parameter 
point $M_{_{Z^{\prime}}} = 200$ GeV and $g_{_{Z^{\prime}}}=0.6$ the same sign top production 
cross section at LHC (only for $u(\bar{u})u(\bar{u}) \to t(\bar{t})t(\bar{t})$ channel) 
is about 50pb which will lead to about 2500 same sign dilepton 
events at 1fb$^{-1}$. Note that the  $t\,t$ production will be much more than the $\bar{t}\,\bar{t}$ 
production because of the difference in the valence and sea quark fluxes in the initial state. Hence 
the same sign dilepton final state will contain more $l^+ l^+$ events than $l^- l^-$ events. New physics 
models like supersymmetry or universal extra dimension also have such same sign dilepton signals but 
generally with similar number of events in the  $l^+ l^+$ and $l^- l^-$ final states.

If we do not see any excess in the production of same sign top pair at Tevatron as well as at LHC, 
then that will conclusively rule out the $Z^{\prime}$ explanation of the Tevatron asymmetry. Still, 
the t channel vector boson exchange as a possible explanation of the Tevatron Asymmetry 
cannot be ruled out by the non-observation of excess same sign top pair events. 
This is because instead of considering a new $Z^{\prime}$ if a new t channel $W^{\prime}$ 
exchange is considered then no such excess is expected. 
Such a model was proposed in \cite{Cheung:2009ch,Cheung:2011qa} with the Lagrangian 
\be
\mathcal{L} \ni -g_{_{W^{\prime}}} \bar{t} \gamma^{\mu}(g_{_L} P_L + g_{_R} P_R) 
d \, W^{+\prime}_{\mu} + h.c. \;.
\ee
Here $g_{_{W^{\prime}}}, g_{_L}, g_{_R}$ are the new coupling constants. 
It was observed in \cite{Cheung:2009ch} that 
the results with only $g_{_L}$ or only $g_{_R}$ are similar but only $g_{_R}$ explains the data in a more 
consistent way\cite{Cheung:2011qa}. We fix $g_{_L}=0,g_{_R} =1$ and take $g_{_{W^{\prime}}}$, 
$M_{_{W^{\prime}}}$ as free parameters. 

\begin{figure}[h]
\centering
\hspace*{0mm}\includegraphics[width=0.75\textwidth]{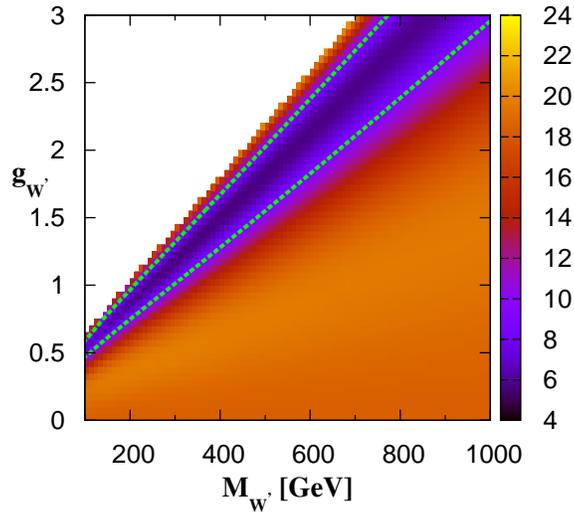} 
\caption{
$\chi^2$ distribution in the $M_{_{W^{\prime}}}- g_{_{W^{\prime}}}$ plane for the 
$W^{\prime}$ model. The area between the two dashed lines corresponds to the 99\% CL region.}
\label{Fig.3}
\end{figure}

In Fig.\ref{Fig.3} we show the $\chi^2$ distribution in the $M_{_{W^{\prime}}}- g_{_{W^{\prime}}}$ 
plane for the $W^{\prime}$ model. The result is similar to the $Z^{\prime}$ case except that for the 
$W^{\prime}$ case slightly larger coupling is required for the same values of the vector boson masses as 
compared to the $Z^{\prime}$ model.

\begin{figure}[h]
\centering
\hspace*{0mm}\includegraphics[width=0.6\textwidth]{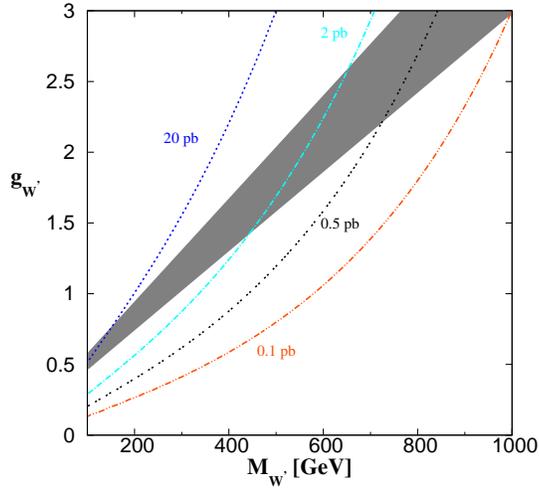}    
\caption{
$W^{\prime}$ $W^{\prime}$ production in the $d \bar{d}\to W^{\prime} W^{\prime}$ channel 
in the $W^{\prime}$ model at 7 TeV LHC. The lines are for constant crosssection contours. 
The shaded region is the 99\% CL allowed region 
of Fig.\ref{Fig.3}.}
\label{Fig.4a}
\end{figure}

\begin{figure}[h]
\centering
\hspace*{0mm}\includegraphics[width=0.6\textwidth]{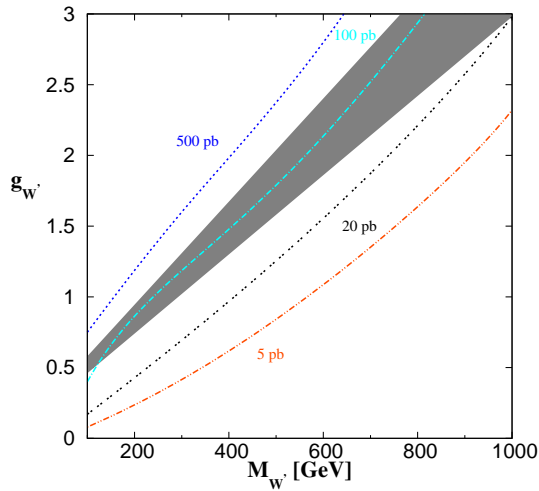}
\caption{
The combined production cross section(fb) of $W^{\prime} t (\bar{t}) $ and 
$ W^{\prime} t(\bar{t}) \bar{d}(d) $ at 7 TeV LHC. The shaded area is the 99\% CL allowed region 
of Fig.\ref{Fig.3}.}
\label{Fig.4b}
\end{figure}

At colliders $W^{\prime}$s can be pair produced via $d \bar{d} \to W^{\prime} W^{\prime}$ or 
can be produced in association with a t quark via $d \, g \to  W^{\prime} t(\bar(t))$(and also 
$d \bar{d} / g g \to W^{\prime} t(\bar{t}) \bar{d}(d) $) channels\cite{Gresham:2011dg}. 
In Fig.\ref{Fig.4a} and Fig.\ref{Fig.4b} we show the production cross sections of $W^{\prime}$ in 
these two channels at the 7 TeV LHC. One can see that the dominant production mode for 
$W^{\prime}$ is the associated production channel. As an example, for 
$M_{_{W^{\prime}}} = 200$ GeV and $g_{_{W^{\prime}}}=0.85$ the $W^{+} W^{-}$ production cross section is 
about 10 pb while the production cross section in the $W^{\prime} t $ and 
$ W^{\prime} t(\bar{t}) \bar{d}(d) $ channels is about 90 pb. For larger 
$W^{\prime}$ masses the cross section decreases rapidly and will be difficult to discover with early 
LHC data.

If $W^{\prime}$ is heavier than 
top quark, it can decay to top quark and will contribute to the $t\bar{t}$ production. Unlike 
$Z^{\prime}$, the $W^{\prime}$ model does not give rise to new channels for the same sign 
top pair production. Hence non observation of excess 
number of same sign top events cannot rule out the $W^{\prime}$ explanation of the Tevatron asymmetry. 
 
LHC, being a $p \, p$ machine, has no directional preference and hence, no 
forward-backward asymmetry can be formed. 
Thus, we focus on another observable called the One sided forward backward asymmetry at LHC 
and study the prediction of the $W^{\prime}$ model for this observable. This will be the content 
of the next section. 
\section{One sided forward backward asymmetry at LHC}
\label{section3}
As mentioned before, unlike Tevatron LHC does not have any preferred direction to produce 
the FB asymmetry and hence the definition of $A_{FB}^{t \bar{t}}$(see Eq.~\ref{eq.1}) is not 
applicable for LHC \cite{Langacker:1984dc,Diener:2009ee}. On the other hand , the momentum distributions of the valence and sea 
quarks inside the proton 
are different. For example, for the subprocess $ d \bar{d} \to t \bar{t}$ very often the $d$ 
quark will have more velocity than the $\bar{d}$ 
quark which gives a non-zero and positive z component of $t \bar{t}$ total momentum in the 
lab frame(i.e., $P^{t\bar{t}}_z >0$). 
Unfortunately, this asymmetry will be erased with the opposite $P^{t\bar{t}}_z$ for the 
subprocess $\bar{d} d \to t \bar{t}$. One way to observe such an asymmetry at the LHC is to put 
a cut on $P^{t\bar{t}}_z$. Note that the gluon contribution is completely symmetric and it is 
the dominant $t\bar{t}$ production channel at the LHC. In order to reduce the gluon contribution 
 one can impose a lower cut on the invariant mass $M^{t\bar{t}}$ of the $t\bar{t}$ system. 
  
Keeping this fact in mind a quantity called one sided forward-backward 
asymmetry was constructed in ref.~\cite{Wang:2010du} which is defined as: 
$$
A_{OFB} \equiv 
\frac{\sigma(\Delta Y > 0)-\sigma(\Delta Y < 0)}{\sigma(\Delta Y > 0)+\sigma
(\Delta Y < 0)}|_{P^{t\bar{t}}_z > P^{cut}_z,M^{t\bar{t}} > M^{cut} } \; .
$$  
Here $P^{t\bar{t}}_z$ is the $z$ component of the total momentum of the 
$t \,\bar{t}$ system in the  $p p$ centre of mass frame.

\begin{figure}[t]
\centering
\includegraphics[width=0.6\textwidth]{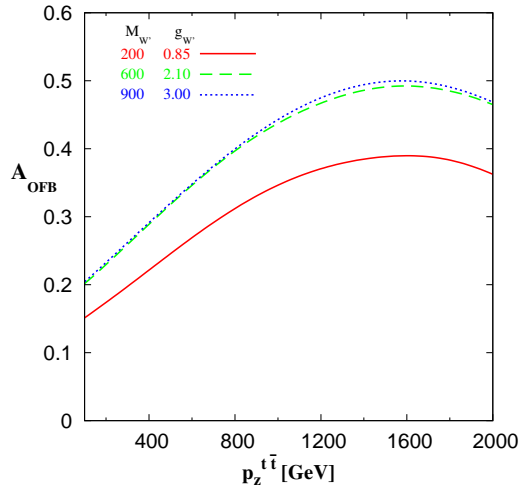} 
\caption{ 
One sided forward backward asymmetry at 7 TeV LHC for the $W^{\prime}$ model.}
\label{Fig.5a}
\end{figure}

\begin{figure}[h]
\centering
\includegraphics[width=0.6\textwidth]{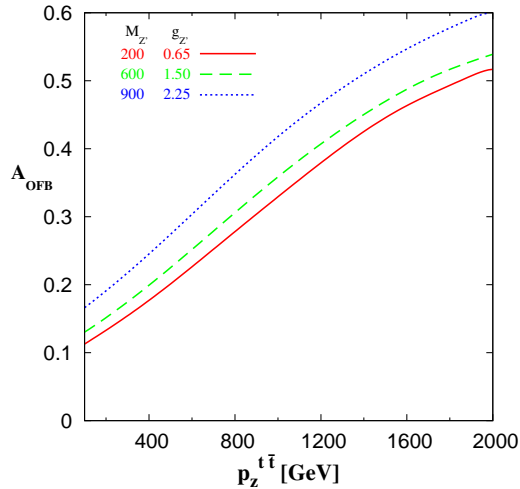}
\caption{
One sided forward backward asymmetry at 7 TeV LHC for the $Z^{\prime}$ models.}
\label{Fig.5b}
\end{figure}

Similar to $A_{FB}$, $A_{OFB}$ also gets contribution at the order $\alpha_s^3$ 
in the SM. For the SM prediction we refer the reader to Fig.5 and 6 of 
\cite{Wang:2010du}. 
We consider three benchmark points (corresponding to low value of $\chi^2$) for the $W^{\prime}$ model  
and calculate the one sided forward backward asymmetry at LHC for 7 TeV center of mass 
energy. We show the results in Fig.~\ref{Fig.5a} as a function of $P^{cut}_z$ for 
$M^{t\bar{t}} > 500 $GeV. 
The size of the asymmetry increases when $M^{t\bar{t}}$ cut is increased. 
We observe that $W^{\prime}$ model predicts quite large $A_{OFB}$ for a centre of 
mass energy of 7 TeV in $pp$ collision and hence, LHC can verify this prediction by measuring $A_{OFB}$. 
Thus, if we do not see any excess same sign top events but observe large one sided 
forward backward asymmetry then that would motivate more detailed study of the $W^{\prime}$ 
model.
Further, large values of the one sided forward backward asymmetry is also possible for the 
$Z^{\prime}$ model (shown in Fig.~\ref{Fig.5b}) and perhaps in 
many other new physics scenarios. 
The shape of the variation of $A_{OFB}$ with $P^{cut}_z$ may 
help to differentiate these models, though, distinguishing many different models conclusively 
will probably require more specific 
signatures and much more detailed studies. We have checked that even for $P^{t\bar{t}}_z > 1500$GeV 
most of the top quarks produced will have rapidity less than 3 and $p_T$ less than 
300 GeV. As a consequence, these top quarks can be detected using conventional techniques and 
more challenging techniques like boosted top algorithms are not required. 
\section{Discussion and Summary}
\label{section4}
We have investigated the possible explanation of the measured FB-asymmetry 
at Tevatron in the framework of $Z^{\prime}$ and $W^{\prime}$ models. 
We perform a $\chi^2$ analysis using the FB-asymmetry measured in different 
rapidity ($\Delta Y$) and $t \bar t$ invariant mass ($M_{t \bar t}$) 
regions. We find that only a small region in the parameter space can 
accommodate the measured cross section and FB-asymmetry simultaneously. 

The $Z^{\prime}$-model predicts production of $t \bar{t}$, 
same sign top pairs and single tops whereas $W^{\prime}$-model predicts 
only production of $t \bar{t}$ at the LHC. Though both these models can 
explain the recent Tevatron measurements, however, we argue that 
non observation of excess of same sign top events may exclude the 
$Z^{\prime}$-model and Tevatron itself has the potential to do so with the 
current data. 

LHC being a pp-machine one does not have the freedom to define the 
FB-asymmetry as defined in case of Tevatron. Thus we study 
the recently proposed one-sided FB-asymmetry that can be measured 
at the LHC. We choose a few benchmark points consistent with Tevatron 
measurements and calculate this asymmetry for both the $Z^{\prime}$ and 
$W^{\prime}$ models. We find the size of this 
asymmetry quite large and can be measured at the LHC even 
running at a center of mass energy of 7 TeV.

To summarize, we investigate the possibility of discriminating 
$Z^{\prime}$ from $W^{\prime}$ model by same sign top quark pair signals and 
measuring one sided forward-backward 
asymmetry at the LHC using the recent measurements of Tevatron as inputs. 
We conclude that non observation of excess of same sign top events and 
observation of large one sided forward-backward asymmetry at the LHC may exclude 
the $Z^{\prime}$ model and point towards a $W^{\prime}$ like scenario.


%
\end{document}